\documentstyle[epsfig]{mn}

\def\spose#1{\hbox to 0pt{#1\hss}}

\def\=#1{\overline{#1}}

\def\lta{\mathrel{\spose{\lower 3pt\hbox{$\mathchar"218$}}
     \raise 2.0pt\hbox{$\mathchar"13C$}}}
\def\gta{\mathrel{\spose{\lower 3pt\hbox{$\mathchar"218$}}
     \raise 2.0pt\hbox{$\mathchar"13E$}}}

\def\kms{{\rm\,km\,s^{-1}}}
\def\kpc{{\rm\,kpc}}
\def\Mpc{{\rm\,Mpc}}
\def\msun{{\rm\,M_\odot}}

\def\pc{{\rm\,pc}}

\def\yr{{\rm\,yr}}

\title[AGN and minor mergers]{Active galactic nuclei and the minor merger hypothesis}
\author[P. Kendall]{Philip~Kendall$^1$, John~Magorrian$^{1,2}$ and
J.E.~Pringle$^1$ \\ $^1$ Institute of Astronomy, Madingley Road,
Cambridge, CB3 0HA, UK \\ $^2$ Department of Physics, University of
Durham, South Road, Durham DH1 3LE \\ }

\date {Accepted ---. Received ---; in original form ---.}
\pagerange{\pageref{firstpage}--\pageref{lastpage}}

\begin{document}
\label{firstpage}
\maketitle

\begin{abstract}

We have investigated the dynamics of the merging process in the minor
merger hypothesis for active galactic nuclei.  We find that for a
satellite galaxy to be able to merge directly with the nucleus of the
host galaxy (for example, to give rise to the compact dust discs which
are seen in early type active galaxies) requires the initial orbit of
the satellite to be well aimed. For the case of the host galaxy being
a disc galaxy, if the initial orbits of the satellites are randomly
oriented with respect to the host galaxy, then the orbits of those
which reach the host nuclear regions in a reasonable time, are also
fairly randomly oriented once they reach the nucleus. We note that
this result might be able to provide an explanation of why the jet
directions in the nuclei of Seyfert galaxies are apparently unrelated
to the plane of the galaxy discs.

\end{abstract}


\section{Introduction}
\label{introduction}

In this paper we investigate the hypothesis that activity in galactic
nuclei is induced by the acquisition and disruption of a small
gas-rich companion galaxy. There has been considerable discussion of
this `minor merger' hypothesis in the literature (see, for example,
recent discussions in Taniguchi, 1999; Chatzichristou, 2000,a,b; 2001
a,b, and papers therein). 

In elliptical galaxies, a study of the effects of minor mergers is
motivated by the observational evidence (van Dokkum \& Franx, 1995)
that dust discs are present in the cores of a high fraction of
ellipticals, with a higher detection rate in radio galaxies (Verdoes
Kleijn et al., 2000). This result is consistent with the results of
surveys of 3CR radio galaxies by de Koff et al. (2000) and by Martel
et al. (1999). The simplest interpretation of these is that the dust
structure seen in these galaxies is the debris trail of a small merger
incident. The analysis of de Koff et al. (2000), and the finding by
Sparks et al. (2000) that galaxies with optical jets tend to display
face-on (i.e. round) dust discs, both suggest that the radio jets in
radio galaxies show a tendency to be aligned perpendicular to the
observed dust disc structure. In this picture a recent merger would
have stimulated the current nuclear activity. The surrounding dust
structure would indicate the plane of the orbit of the disrupted
satellite, and the resulting radio jets would naturally be expected to
be perpendicular to this plane.

In view of this, we investigate first the conditions required for a
small gas-rich satellite galaxy to be accreted by a larger galaxy in
such a way that it gives rise the a dust disc similar to those
observed. To focus the discussion we concentrate on modelling the
formation of the dust disc in the otherwise unremarkable E1 galaxy
NGC~3379. Our results are straightforwardly applicable to other
gas-poor ellipticals. The dust disc in NGC~3379 (van Dokkum \& Franx,
1995; Ferrari et al., 1999) has a radius $\sim 100\pc$, and a mass
estimated to be around $150\msun$. In addition Ferrari et
al. (1999) show that there is more patchy dust extending out to about
$1\kpc$ from the nucleus. We adopt a simplified procedure to model
the interaction. We use spherical models for both the major galaxy and
the minor mergee and model the motion of the orbit of the satellite
using simple dynamical friction, along with a straightforward model
for the tidal disruption of the satellite (Section 2). We use a simple
model for the subsequent evolution of the stripped gas (Section
3). Our basic conclusion (Sections 5.1 and 6) is that in order to
generate a dust disc of the kind observed it is necessary for the
initial trajectory of the merging satellite to be accurately directed
towards the nucleus of the major galaxy.

However, reality turns out to be more complicated than this simple
picture might suggest. First there is the recent analysis by Schmitt
et al. (2002) who consider only those E/S0 galaxies with active nuclei
which have ${\it well-defined}$ dust discs in their central regions,
rather than galaxies whose dust distribution is more
dispersed. Schmitt et al. conclude that the radio jets from the
centres of these discs are not in fact oriented perpendicular to the
dust discs' axes, and the angles the jets make with the axes are
consistent with being distributed evenly in the range $0\degr -
75\degr$. Second there is the finding by Kinney et al (2000) that in
Seyfert galaxies the directions of the central jets are consistent
with being oriented randomly in space, independent of the plane of the
gas disc of the host spiral galaxy. If the minor merger hypothesis is
to hold, then the apparent misalignments between between the angular
momentum of the accreting material (as evidenced by the observed discs
in early type galaxies and by the disc plane in Seyferts) and the
observed direction of the jets needs some explanation. One possible
explanation is the idea that the jet direction is governed by the spin
of the central black hole (through the Bardeen-Petterson effect
(e.g. Scheuer and Feiler, 1996 and references therein) and that for
some reason the spin of the black hole is misaligned with the current
accretion flow. While this is a reasonable explanation if the galaxy
is spherically symmetric it would seem difficult to achieve if the
spin of the black hole has been generated by the gas accretion process
and if the potential of the galaxy is such the the accreted gas would
(in the long term) be expected to have a preferred direction for its
angular momentum vector. One way round this problem, following the
ideas of Wilson \& Colbert (1995; see also Merritt, 2002), is if the
black hole spin is generated not by accretion of gas, but rather by
accretion of black holes presumably contained in the nuclei of the
small merging satellites. Even a small black hole (say a tenth of the
mass of that in the host galaxy) can in its final approach add
substantial orbital angular momentum to the spin of the final
coalesced black hole. For this process to work, what is required is
that the central (nuclear) black hole of a merging satellite be able
to reach the nuclear regions of the major galaxy with the angle
between its orbital angular momentum and the axis of symmetry of the
gravitational potential of the major galaxy distributed over a wide
range of angles.

To investigate this possibility we extend our analysis to the case
where the major host galaxy is a spiral galaxy. Here we are not
concerned with the fate of any gas that might have been in the merging
satellite, but are, rather, concerned with the fate of its nuclear
black hole. We make use of similar simplified dynamics using the
concepts of gravitational drag and a similar approximation to take
account of tidal stripping to those we used before, but we now apply
these concepts to a a galaxy with a more complicated internal
structure (Section 4). As before we find that the cross-section for
an incoming satellite to reach the nucleus of the host galaxy in a
Hubble time is relatively low. However here we pose the additional
question: of those satellite nuclei which do reach the central regions
of the host galaxy in a reasonable time, what is the distribution of
the inclinations of their orbits relative to normal to the disc of the
galaxy? In order for the merger hypothesis to remain viable we shall
require the distribution we find to very broad. We summarise our
findings and give discussion of our results in Sections 5.2 and 6
respectively.

\section{Merging with a spherical galaxy}

In this Section we describe our simplified galaxy dynamics and
merging procedure.

\subsection{Galactic models}
\label{galacticmodels}

As mentioned in Section 1, we first investigate the particular case in
which the major galaxy is spherically symmetric as a reasonable
approximate model for the merging of a satellite with the galaxy NGC
3379.

\subsubsection{Major galaxy (NGC 3379)}

To model the density profile of NGC~3379 we have made use of Lima
Neto, Gerbal and M\'arquez's \shortcite{limaneto99} analytic
approximation to the deprojection of its surface brightness profile
obtained with HST \cite{pastoriza00}\footnote{Note that there is a
typographical error in Pastoriza et al.\ (2000): in equation 2 `exp'
should be `10 to the power of' (Ferrari, private communication)}. Lima
Neto et al. \shortcite{limaneto99} assume that the galaxy is
spherical, with a constant mass-to-light ratio $\Upsilon$, and find
that
\begin{equation}
\rho(r) = \rho_0 \left( \frac{r}{a} \right) ^ {-p}
\exp \left[ - \left( \frac{r}{a} \right) ^ \nu \right],
\end{equation}
where $\nu = 0.424$ and $p = 0.572$. Using the normalization to the
surface brightness given by Pastoriza et al. \shortcite{pastoriza00},
together with an assumed distance to NGC~3379 of $9.9 \Mpc$ and a mass
to light ratio of $\Upsilon = 5.3\Upsilon_\odot$ (Magorrian et al.\
1998), gives $\rho_{\rm{o}} = 3.96 \times 10^3 \,
\rm{M}_\odot/\rm{pc}^3$, and $a = 18.6\pc$.  The resulting visible
mass is $2.9 \times 10^{10}\msun$, with a half-mass radius of
about $800\pc$, excluding any dark halo.  We find that the inclusion
of a dark halo or a central $10^8\msun$ black hole (Magorrian et
al.\ 1998; Gebhardt et al.\ 2000) has negligible effect on our results
below.

We calculate the mass distribution, potential and velocity dispersion
of the galaxy through the usual equations
\begin{eqnarray}
M(r) & = & \int_0^r \rho(r') 4 \pi r'^2 \textrm{d} r' \\
\Phi(r) & = & - \int_r^\infty \frac{G M(r')}{r'^2} \textrm{d} r' \\
\sigma^2(r) & = & \frac{1}{\rho(r)} \int_r^\infty \frac{G M(r')
\rho(r')}{r'^2} \textrm{d} r',
\end{eqnarray}
where we have assumed for simplicity that the galaxy's velocity
dispersion tensor is isotropic.  We evaluate the integrals adaptively
using Simpson's rule until an accuracy of one part in $10^6$ is
reached and store the calculated values of $M(r)$, $\Phi(r)$ and
$\sigma(r)$ on grids with 30 points spaced logarithmically in radius
between $100 \pc$ and $100 \kpc$.  Values between tabulated points are
obtained via cubic spline interpolation.

\subsubsection{Satellite}
\label{satellitemodel}

We have considered two different models for the small infalling
satellite galaxy.  One is a rigid Plummer sphere with density
distribution given by
\begin{equation}
\rho(r)=\frac{3M}{4 \pi b^3} \left( 1 + \frac{r^2}{b^2} \right) ^ {5/2},
\end{equation}
and the other is a rigid Hernquist sphere with density distribution
given by
\begin{equation}
\rho(r) = \frac{M}{2 \pi} \frac{a}{r} \frac{1}{ \left( a + r \right) ^ 3 }.
\end{equation}
For both of these cases we calculate the mass and velocity dispersion
as for the host galaxy, except with a grid spaced logarithmically
between $0.1 \kpc$ and $10 \kpc$.  The total mass ($M = 10^9 \ \msun$)
and scale lengths ($b = 0.4\kpc$ and $a = 0.23\kpc$) were chosen to
match the estimated mass and half-mass radius ($0.55 \kpc$) of
Sagittarius \cite{helmi00}, as a typical dwarf galaxy.  We will see in
Section \ref{tidalstripping} that the choice of model for the
satellite does not make a significant difference to our results.

\subsection{The orbit of the infalling satellite}
\label{infall}

In computing the orbit of the infalling satellite, we neglect changes
which might be induced by the satellite in the host galaxy. This
approximation has been the subject of debate in the literature
\cite{zaritsky88,hernquist89}, with the conclusion being reached that
it is reasonable to ignore the galaxy response for low mass satellites
\cite{velazquez99}. Physically, this is what is expected, as the
response induced in the galaxy is proportional to the mass ratio of
the satellite to the galaxy ${M_s}/{M_g}$, and the effect of this
response is again proportional to the mass ratio, so the overall
effect will be proportional to $\left({M_s}/{M_g}\right)^2$.

We model the motion of the satellite using Chandrasekhar's dynamical
friction formula \cite{binney87}. Thus
\begin{equation}
\label{chandra}
\frac{\textrm{d}\bmath{v_s}}{\textrm{d}t} = - \frac{k M_s \rho
}{\left|\bmath{v_s}\right|^3}\bmath{v_s},
\end{equation}
where $\bmath{v_s}$ is the velocity of the satellite, $\rho$ is the
density of the medium, $k$ is given by
\begin{equation}
\label{chandrak}
k = 4 \pi \ln \Lambda G^2 \left( \textrm{erf}(X) - \frac{2 X}{\sqrt
\pi} \exp \left( - X^2 \right) \right),
\end{equation}
$G$ is Newton's gravitational constant, $X$, the dimensionless
speed of the satellite, is given by
\begin{equation}
X = \frac{\left|\bmath{v_s}\right|}{\sqrt 2 \sigma(r)},
\end{equation}
and $\Lambda$ is the Coulomb logarithm discussed below.

Equation~(\ref{chandra}) applies to the case of a point mass moving
through an infinite homogeneous medium with a Maxwellian velocity
distribution. However, numerical experiments \cite{velazquez99} show
that it also provides a reasonable approximation to the gravitational
drag experienced in more realistic situations, provided one chooses
the Columb logarithm~$\ln\Lambda$ appropriately.  To determine whether
it is adequate for the situation we are considering, we use it to
reproduce the results from the infall of a satellite into a dark halo
when the galaxy was modelled using a full $N$-body treatment
\cite{vandenbosch99}.  We note that this procedure has been
investigated in some detail by Taylor \& Babul (2001), who demonstrate
that it does provide a reasonable approximation.

To perform the integration, we used the algorithm from the fifth-order
Cash-Karp Runge-Kutta integrator `\texttt{odeint}' \cite{numrec}. The
timestep used is automatically adjusted by the routine to control
errors, and in no case did the timestep reach a significant fraction
of the dynamical time of the satellite.  

\begin{figure}
\centerline{\epsfig{file=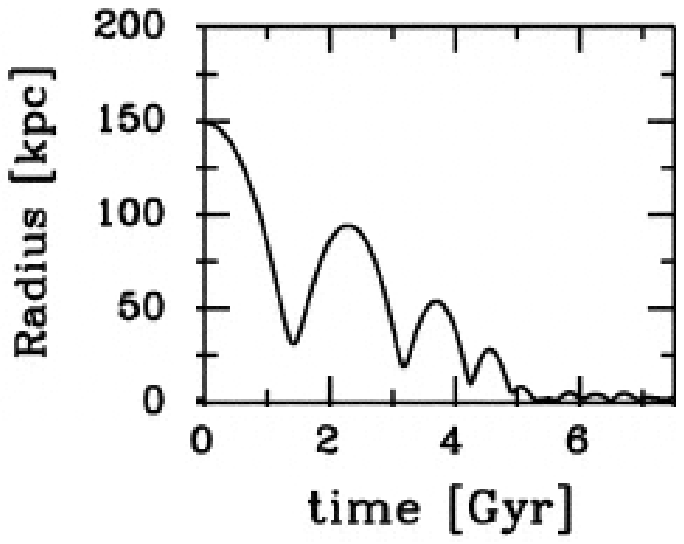,width=0.4\hsize}
            \raise-2pt\hbox{\epsfig{file=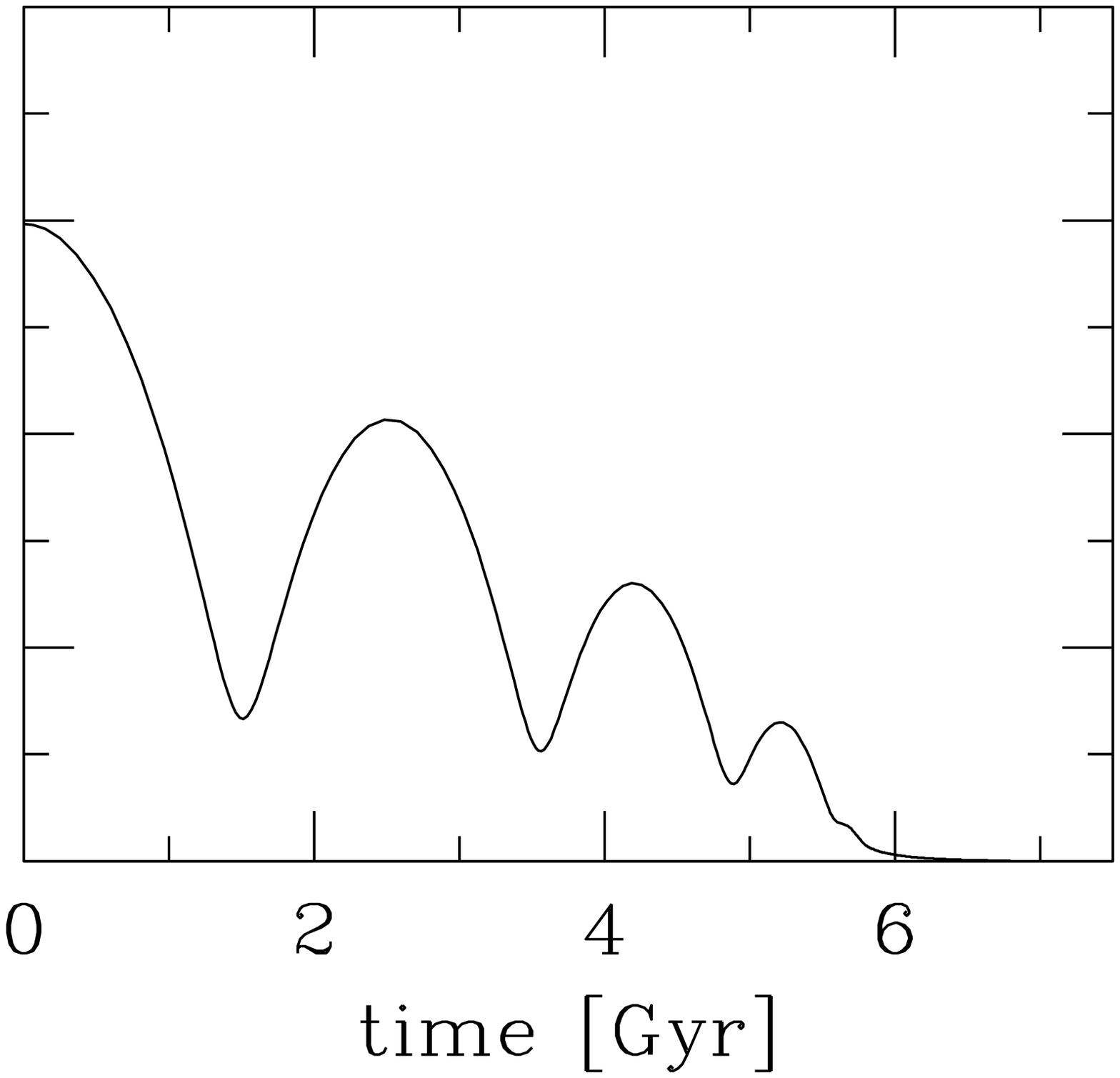,height=29mm,width=31mm}}}
\centerline{\epsfig{file=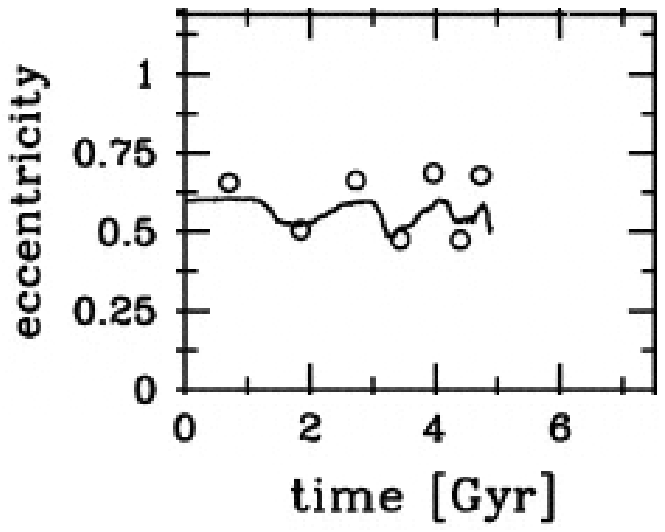,width=0.4\hsize}
            \raise-2pt\hbox{\epsfig{file=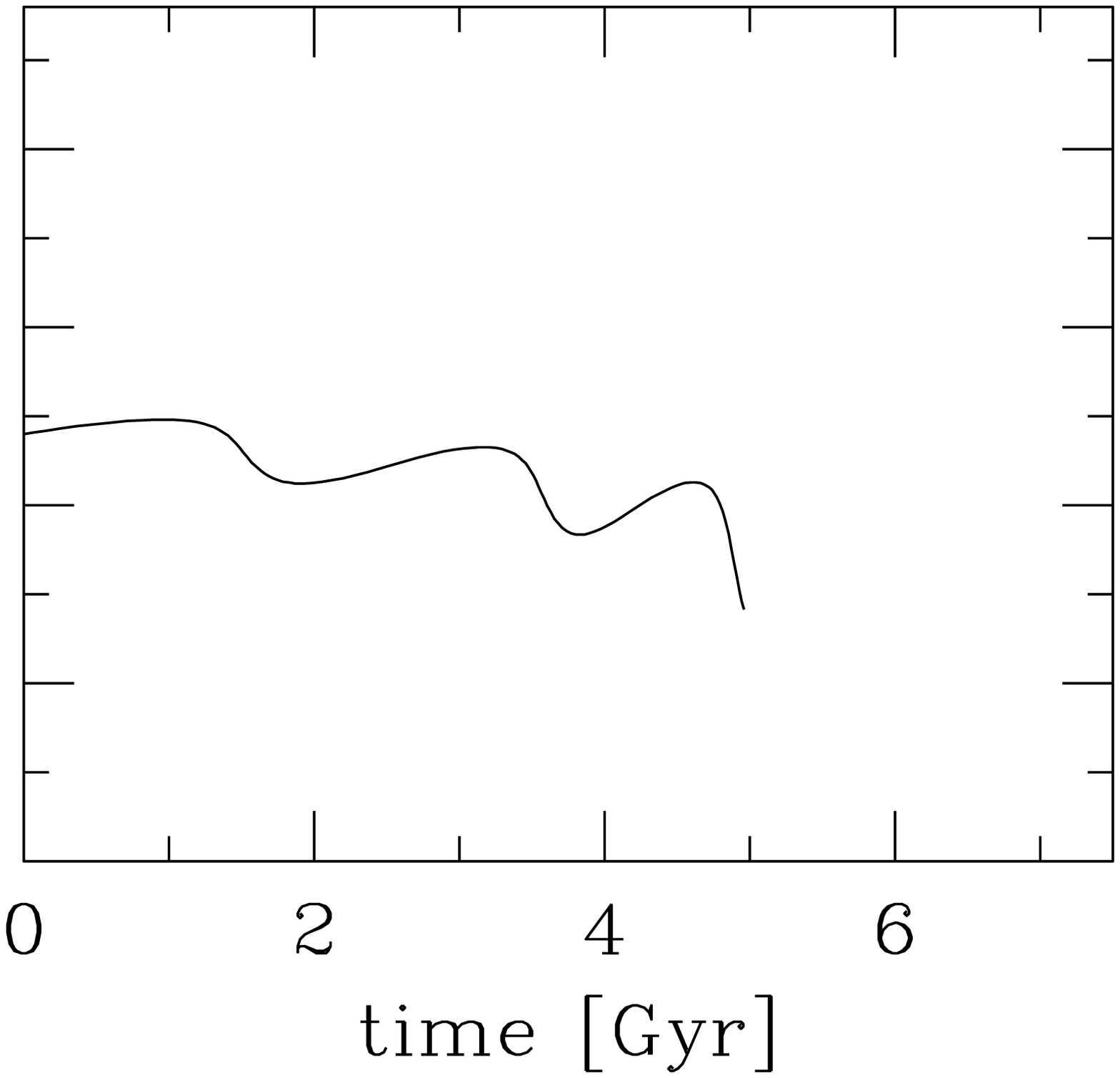,height=29mm,width=31mm}}}
\caption{A comparison our approximate method for following the effects
of dynamical friction on a satellite's orbit (right) against results
from a full $N$-body simulation (van den Bosch et al.~1999; left
panels).  The top panels show how the distance of the satellite from
the centre of the host varies with time.  The bottom two show the
eccentricity of the satellite's orbit, defined as $e\equiv
(r_+-r_-)/(r_++r_-)$, where $r_+$ and~$r_-$ are the satellite's apo-
and peri-centre radii respectively.}
\label{fig-vandenbosch}
\end{figure}

As a test of our procedure, we use our method to model van den Bosch
et al.'s \shortcite{vandenbosch99} Plummer sphere sinking into a
truncated isothermal halo. We find that we are able to reproduce their
results over a wide range of conditions if we choose $\ln
\Lambda\approx2$.  In Figure~\ref{fig-vandenbosch} we show a
comparison between the full $N$-body calculations of van den Bosch et
al (1999) and our approximate procedure for a typical case.  As can be
seen, there is reasonable agreement between the two. Using these
results we shall adopt the same choice $\ln\Lambda=2$ for the orbital
computations contained in this paper.  We should note that, for the
spherical case, our treatment of dynamical friction also ignores the
effects of possible galaxy rotation and velocity anisotropy, but
unless these are extreme we expect that this approximation will be
adequate in the current case.

\subsection{Tidal stripping}
\label{tidalstripping}

In the model used by van den Bosch et al. \shortcite{vandenbosch99},
the satellite maintained a constant mass all the way through the
simulation. Due to tidal effects, the outer regions of the satellite
will not remain bound to the satellite. We are interested, therefore,
in the manner in which the infalling satellite is tidally stripped as
it passes through the host galaxy. This has an effect upon the orbital
dynamics, because the deceleration due to Chandrasekhar friction is
proportional to the mass of the satellite.

We use a simple algorithm to simulate the tidal stripping of the
satellite. At each instant, we define the tidal radius, $r_t$, of the
satellite as being the radius at which the mean density of the
satellite is equal to the mean density of the galaxy, viz.,
\begin{equation}
\label{tidalradius}
\frac{M_s(r_t)}{\frac{4}{3} \pi r_t^3} = \frac{M_g(r_s)}{\frac{4}{3}
\pi r_s^3},
\end{equation}
where $M_s(r)$ is the mass of the satellite contained within a radius
$r$, $M_g(r)$ is the same for the galaxy and $r_s$ is the
galactocentric radius of the centre of the satellite.  We then
truncate the satellite at $r=r_t$, keeping the density structure
internal to this point unchanged. This reduces the mass of the
satellite to $M_s(r_t)$. In reality the satellite will react to this
density truncation on an internal dynamical timescale, but since by
construction\footnote{The tidal radius is defined by equating the mean
density of the satellite to the mean density of the main galaxy, and
dynamical timescales just depend on mean densities.} this timescale is
similar to the orbital timescale at this point, neglect of this effect
is likely to be a reasonable approximation for current purposes.

\begin{figure}
\centerline{\epsfig{file=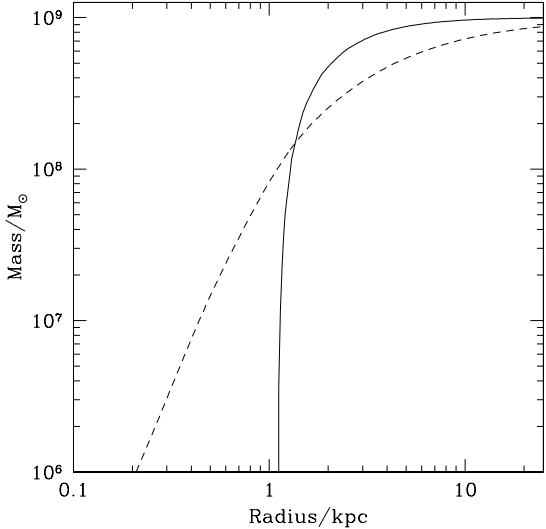,width=0.9\hsize}}
\caption{Remaining satellite mass as a function of distance from the
centre of the host galaxy obtained using the tidal stripping
prescription~(\ref{tidalradius}).  The solid and dashed curves plot the
results for the Plummer and Hernquist models for the satellite
 respectively (section~\ref{satellitemodel}).  The former cuts off at
around $1\kpc$ as its finite central density means that it is
completely disrupted by that point. }
\label{strip}
\end{figure}

In Figure \ref{strip}, we show the outcome of this stripping algorithm
by plotting the fraction of the satellite galaxy which remains bound
as a function of its distance from the centre of the host galaxy. We
show the results for both models of the satellite galaxy. The Plummer
sphere is completely disrupted by the time it comes within 1 kpc of
the centre of the host galaxy. The Hernquist sphere loses 99\% of its
initial mass by the time it comes within is $400\pc$ from the
centre of the galaxy. These distances are much greater than the size
of the observed dust disc, which has a radius of around $100\pc$, but
are not that dissimilar to the overall size of the dust spread within
the galaxy. We note that these result depends only upon the density
profiles of the satellite and the galaxy. They are just the result of
solving Equation~(\ref{tidalradius}), and do not involve any
consideration of the dynamics.

\section{Deposition and location of the stripped gas}
\label{sticky}

\subsection{Low energy orbits}

To make a preliminary investigation of the initial conditions required
for a merging satellite to be able to deposit significant gas into the
nucleus of the host, we start by considering a series of low-energy
orbits for the satellite galaxy. In each one the satellite starts on
an orbit at an apocentre of $10\kpc$ from the galactic centre (that
is, it starts with zero radial velocity at that point), and we vary
the initial varying angular momentum.  We characterize the initial
angular momentum in terms of an eccentricity $e$, where $e =
(r_+-r_-)/(r_++r_-)$ and $r_+(r_-)$ are the apo-(peri-)centres of the
orbit of a test particle with the same initial location and
velocities. We use the models of NGC~3779 and the satellite as
described in Section 2.

To simplify the discussion we shall assume that the gas in the
satellite galaxy is distributed spatially in the same way as the
stellar matter.  We make this assumption for two reasons: First, such
an assumption probably overestimates the concentration of the gas
within the dwarf.  This will lead to the gas being stripped stripped
from the dwarf later, and thus at a lower angular momentum than would
otherwise occur. Hence this assumption will be an upper bound on the
magnitude of any effect seen.  As we show later, even with this
assumption, we cannot replicate the observed effects. Second, by
assuming a spherical dwarf, we can dramatically reduce the
computational power required for our simulations and thus investigate
a larger region of orbital parameter space.

\begin{figure*}
\centerline{
\epsfig{file=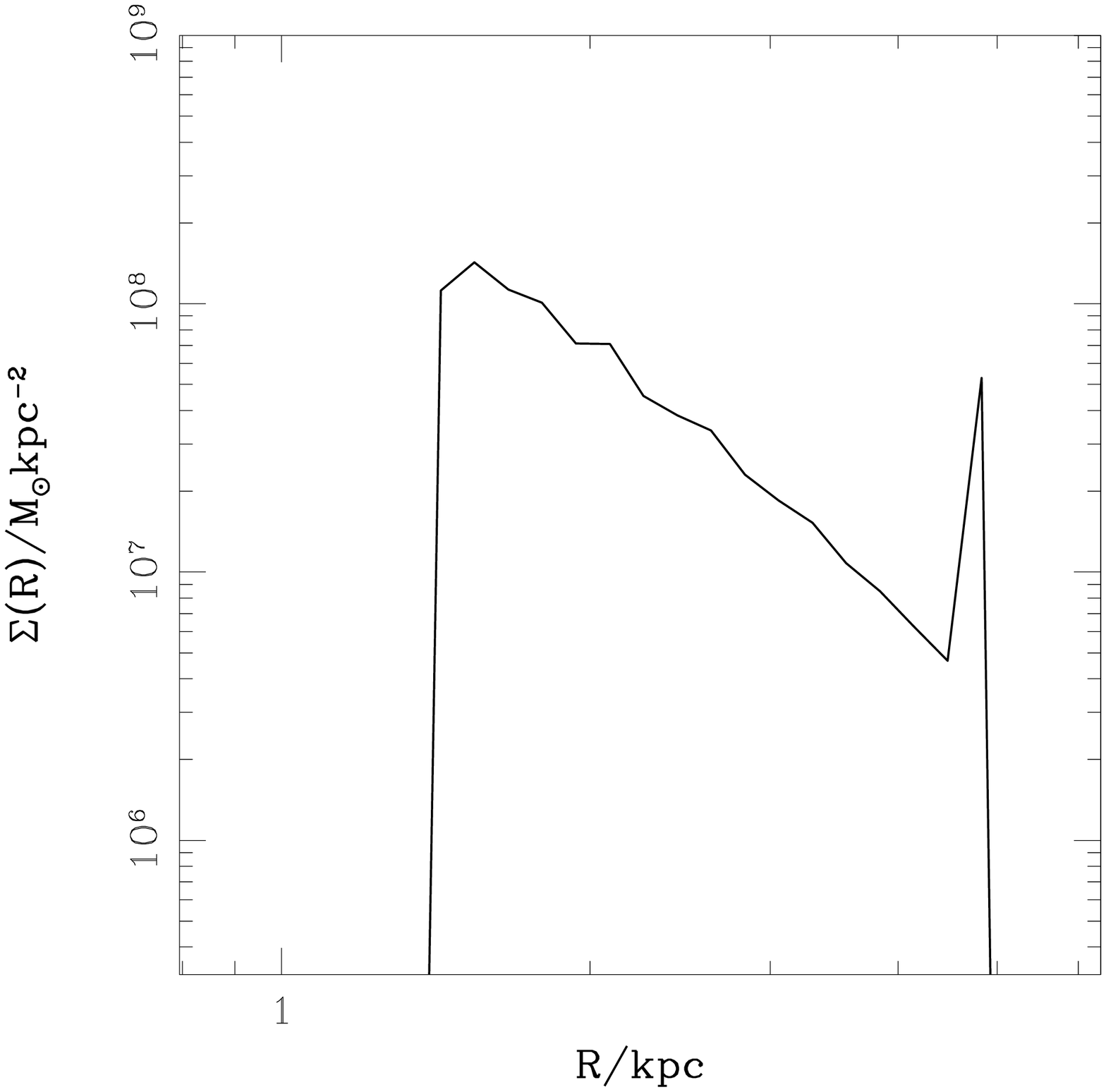,width=0.3\hsize}
\epsfig{file=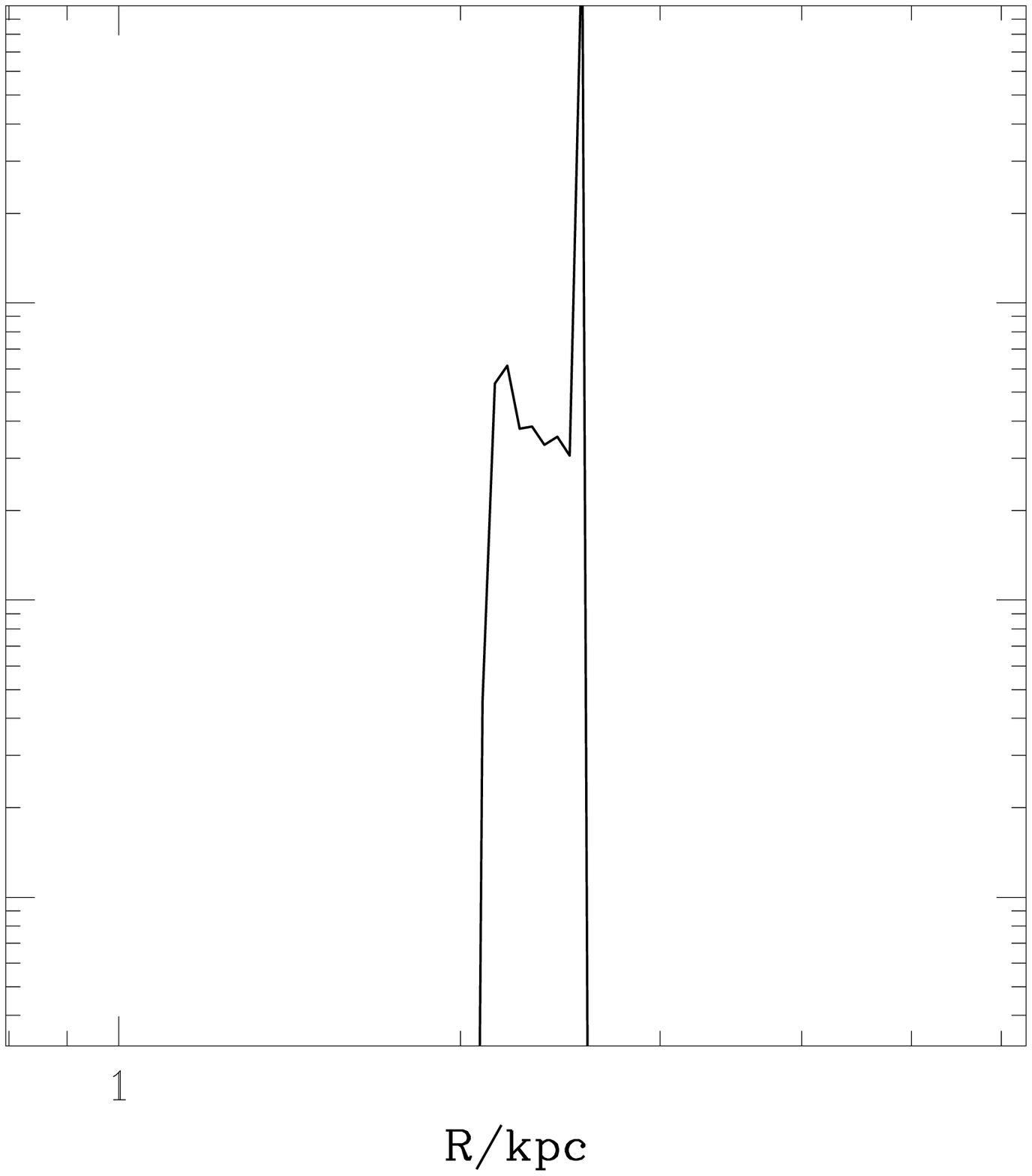,width=0.3\hsize}
}
\caption{The surface density distribution of stripped gas if the gas
is assumed to maintain its angular momentum, but loses energy until
its orbit becomes circular.  The panels show results for initial
eccentricities $e = 0.5$ (left) and $0.75$ (right).  The large spike
in every plot is due to the large amount of stripping which occurs
during the first pericentre passage, whilst the spiky structure seen
in the ring profiles is due to stripping occurring near successive
pericentre passages.}
\label{circularorbits}
\end{figure*}

As an initial estimate of where the stripped gas ends up, the simplest
assumption is that stripped material conserves its angular momentum
about the centre of the host galaxy (since the host is spherically
symmetric), and that each element of gas, by losing energy through
dissipative processes ends up in a circular orbit corresponding to its
initial angular momentum~$L$. The radius~$r_c$ of this orbit is given
by
\begin{equation}
L^2 = G M_g(r_c) r_c.
\end{equation}
The results of making this simple approximation are shown in Figure
\ref{circularorbits}. The outermost spike in each of these mass
distributions is caused by the stripping which occurs at the first
pericentre passage. For a high angular momentum orbit ($e=0.5$), the
stripping occurs over a range of radii as the satellite's orbit
gradually decreases in size due to gravitational drag. For the
intermediate angular momentum orbit ($e=0.75$) the stripping occurs
more rapidly and thus the range of angular momentum in the stripped
gas is reduced.

In order to acheive the required radial penetration, we also
investigated a low angular momentum orbit ($e=0.98$) for which the
satellite's orbit is almost radial.  In this case, the gravitational
drag plays almost no role, and the satellite is totally disrupted on
the first periastron passage, but the gas still ends up at radii
exceeding $250\pc$. Thus we conclude that, using these approximations,
the gas cannot end up within around $100\pc$ unless the satellite was
initially on a very eccentric, indeed essentially radial, orbit.

\subsection{Gas dynamics}

The assumption we made above, that an element of stripped gas, as it
falls in, loses energy by interactions with other elements, but does
not exchange angular momentum with them is not fully realistic. Since
the angular momentum is crucial in determining the ultimate location
of the gas, and since the discs we are obtaining are still much larger
than observed discs we now consider a more realistic model for the
radial evolution of the stripped gas.

In the spirit of the approximations made in this paper, rather than
undertaking a full hydrodynamical treatment (for example using SPH),
we opt for computational simplicity and have followed the approach of
Lin and Pringle \shortcite{lin76}. In such a scheme, the gas particles
move freely in the combined potential of the host galaxy and
satellite, except that a form of dissipation is introduced which
simulates thermally radiative shocks, by preventing the crossing of
particle orbits while exactly conserving mass and angular momentum. In
this scheme, if two particles are close to each other, they interact,
losing a fraction of their energy, but retaining their total
momentum. One effect of this is to redistribute the angular momentum
between different particles. For example, a circular ring of particles
acts as an accretion disc and gradually spreads out to form a wider
ring. In this paper, we consider inelastic collisions between
particles in order to produce the greatest dissipation. This is
clearly the most extreme situation, and therefore the results of this
section will provide an upper bound on the magnitude of this effect.

The scheme used to determine when collisions occur is simple: we
divide the space in the host galaxy into cells, and, after every time
step of the integration, if two (or more) particles occupy the same
cell, they interact.  We use a cylindrical polar grid $(R,\phi,z)$ to
describe the cells, with the $z=0$ plane aligned with the orbit of the
satellite galaxy. The cell boundaries are spaced logarithmically in
$R$, from $100\pc$ to $20\kpc$ with $R_{n+1}/R_n=1.02$, and linearly
in the $z$ from $z=-5.05\kpc$ to $z=5.05\kpc$ with
$z_{n+1}-z_n=0.1\kpc$.  The choice of $R_{n+1}/R_n$ is governed by a
balance between the need for computational efficiency (if the zones
are too small then there are two many cells with too few particles in
them which permits interpenetration of fluid flows), and a need to
keep the zones small, so that viscous effects do not become too rapid,
resulting in unphysically rapid spreading of the gas discs which
form. We performed runs with $R_{n+1}/R_n$ of $1.05$, $1.02$ and
$1.01$. The results from the runs at $1.02$ and $1.01$ were very
similar, implying that convergence had been achieved. Hence we choose
$R_{n+1}/R_n=1.02$ to reduce the computational time required for our
simulations. We choose 314 equispaced points in $\phi$.  This means
that the projection of the cells onto the $z=0$ plane is approximately
square and that the effective cross section of our gas particles
varies from $2\pc$ at $R=100\pc$ to $400\pc$ at $R=20\kpc$.

We create one gas particle, of mass $10^5\msun$ for every $10^5
\msun$ stripped from the satellite.  The particle is started with a
velocity
\begin{equation}
\bmath{v_g} = {\mathbf v}_s + \sigma_s\left( \left|\bmath{r_g} -
\bmath{r_s}\right| \right)\bmath{\hat{r}},
\end{equation}
where $\bmath{v_g}$ and $\bmath{v_s}$ are the velocity of the gas
particle and the satellite respectively, $\bmath{r_g}$ and $\bmath{
r_s}$ are the positions of the centres of the gas particle and the
satellite, $\bmath{\hat{r}}$ is a unit vector in a random direction
and $\sigma_s(r)$ is the velocity dispersion of the satellite at a
radius $r$. A particle is removed from the computation when one of the
following conditions is met: the particle is within $100 \pc$ of the
centre of the galaxy (at which point, the gas is considered to have
become part of the nuclear gas disc); the particle is within 90\% of
the current tidal radius of the satellite (when it is considered to
have been reabsorbed by the satellite); or, the particle moves beyond
the outermost edges of the grid at $ R = 20 \ \kpc$ and $\left| z
\right| = 5.05 \ \kpc$.

\begin{figure*}
\centerline{
\epsfig{file=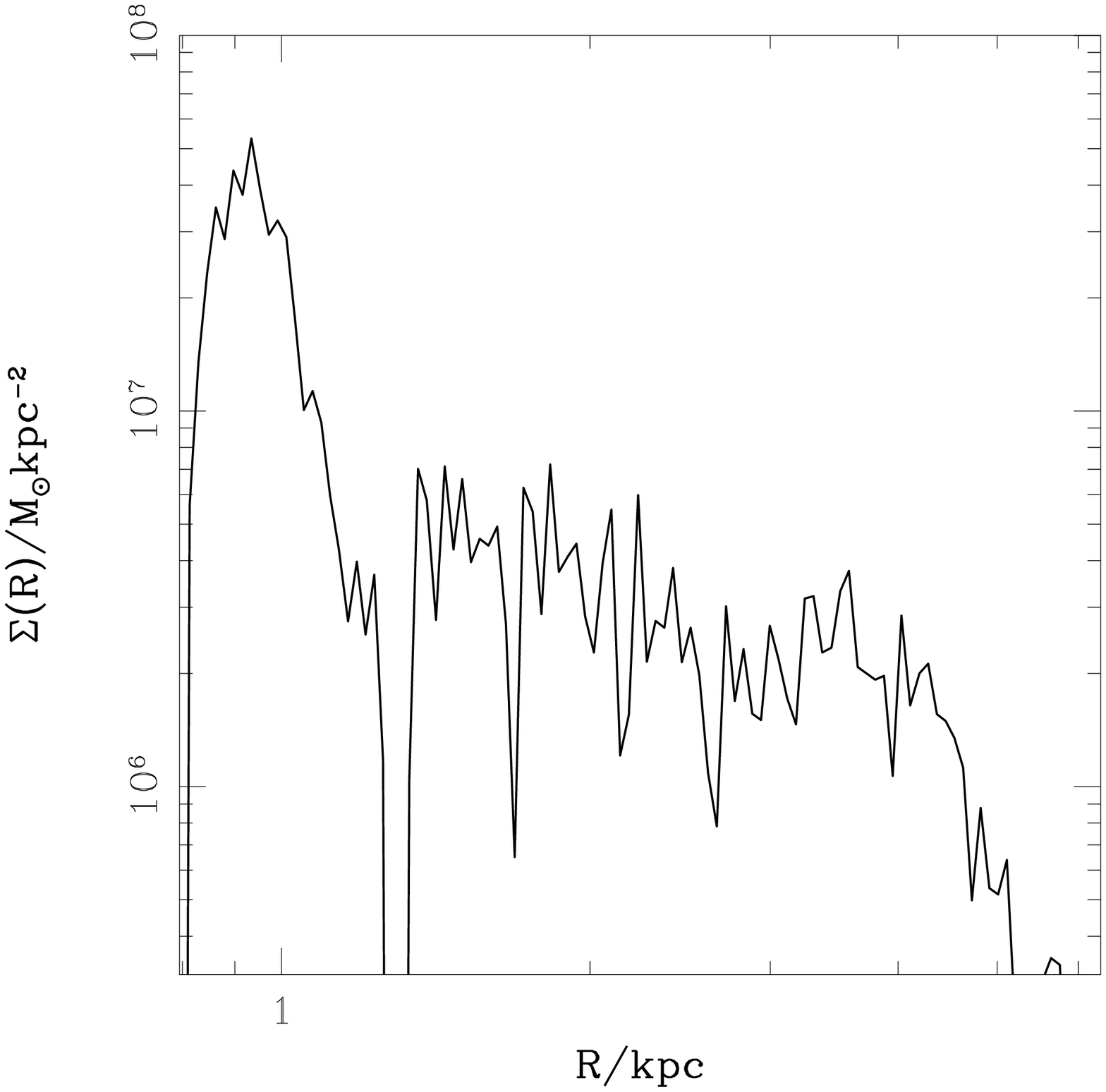,width=0.3\hsize}
\epsfig{file=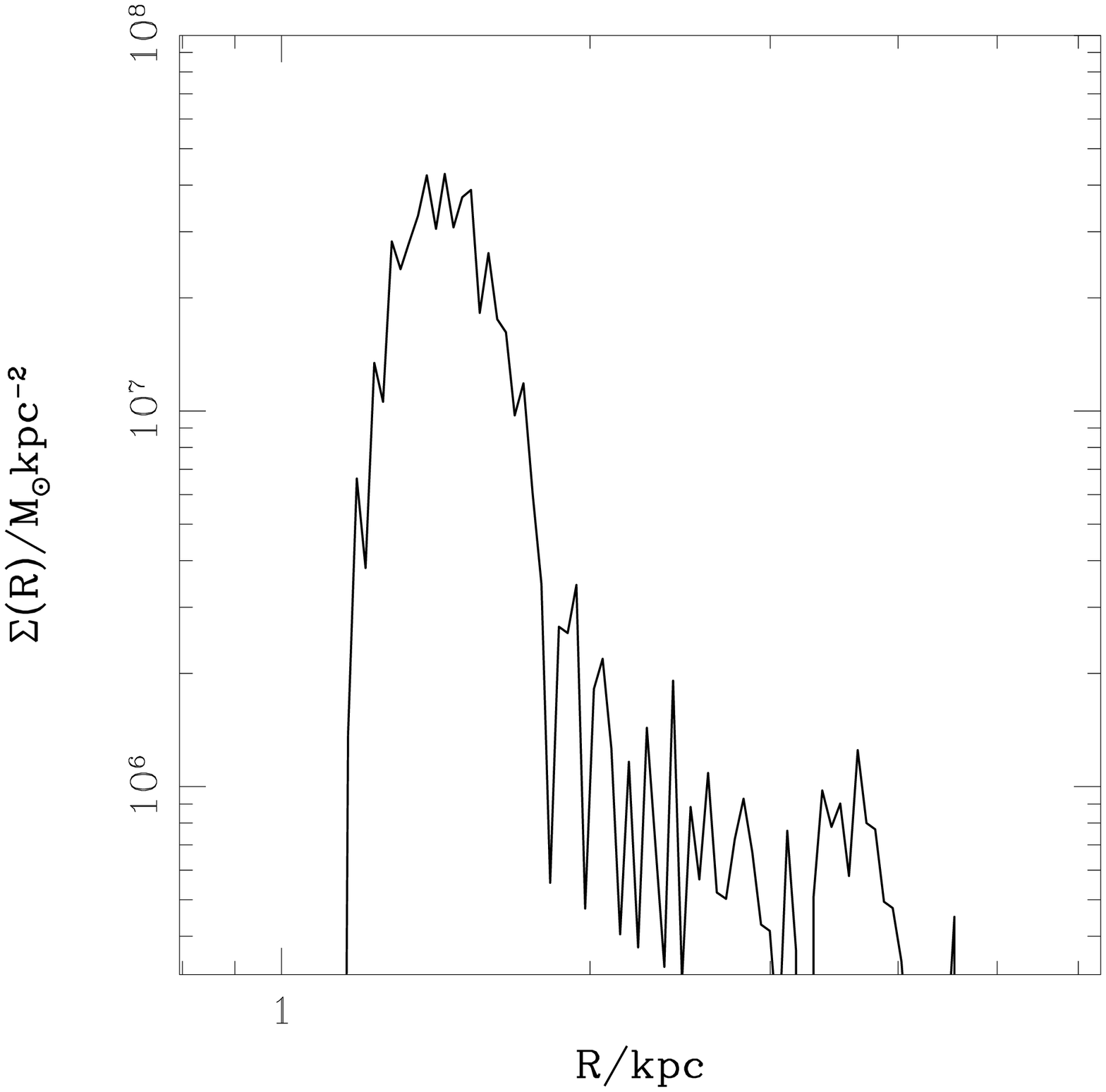,width=0.3\hsize}
}
\caption{The surface density distribution of gas stripped according to
the model in Section~\ref{sticky} for orbits with initial
eccentricities of $0.5$ (left panel) and 0.75 (right panel).  Results
for an eccentricity of $0.98$ are not shown as the majority of
the mass is removed from the simulation; see Section
3.2 for details.}
\label{rings}
\end{figure*}

The radii at which the gas settles initially using this model are
shown in Fig.~\ref{rings}. For comparison with the previous estimates
(Figure~\ref{circularorbits}) we show the result for initial
eccentricities of $0.5$ and $0.75$.  We do not show the results for an
initial eccentricity of $0.98$, because the majority of the stripped
mass passes within $100\pc$ of the galaxy centre and is therefore
removed from the simulation.  By comparing these profiles with those
in Figure \ref{circularorbits}, it can be seen that the result of
allowing the gas particles to interact with each other in a more
realistic manner is to broaden the distributions, and to move the
innermost gas particles inwards by about a factor of two in
radius. Some of this difference is due to the more realistic treatment
of the stripping process, some is due to treating the gas dynamics
more realistically and some is due to the artifical viscous effects
introduced by our approximate scheme for treating the gas dynamics.
We believe that the viscous effects are not significant here.  The
mean radius of the final gas distribution is within 20\% of the
results obtained using the simple prescription of the previous
section, which implies the total energy loss is similar. This
approximate similarity between the two approaches serves to indicate
that any viscous effects introduced due by our grid-based dissipation
scheme are reasonably small.

\subsection{High energy orbits}
\label{higheccentricity}

The above results make it clear that in order to form a dust disc of
the size of a few hundred parsecs of the kind which is seen in
NGC~3379, the orbit of the hypothesized incoming satellite galaxy is
not of the form of a low energy, low eccentricity orbit which goes
through multiple pericentre passages before the satellite is
disrupted. Rather the initial orbit of the incoming satellite needs to
have a sufficiently high eccentricity (i.e. low angular momentum) that
the satellite is likely to be completely disrupted on its first pass.

For a more detailed investigation of the high eccentricity regime, it
is more useful to parameterize the problem in terms of an impact
parameter $b$ and an initial velocity $V_0$. In addition, to model the
approach of the satellite galaxy to the host's core, we now start the
satellite's orbit at a galactocentric radius of $80 \kpc$. We use the
same procedure as outlined above (Section 3.2) to treat the gas
stripping and subsequent gas dynamics, but, since the pericentric
distances are now considerably reduced, we now decrease the inner
radius at which particles are removed from the computation from $100$
to $10\pc$.  This is necessary because of the change in the nature of
the orbits of the gas particles.  When we were considering low energy
orbits, the particles which reached $100\pc$ were on essentially
circular orbits and are assumed to become part of the nuclear gas
disc, but in the high eccentricity case, particles reach $100\pc$ with
velocities much greater than circular and as such do not necessarily
get incorporated into the disc.

\begin{figure*}
\centerline{
\epsfig{file=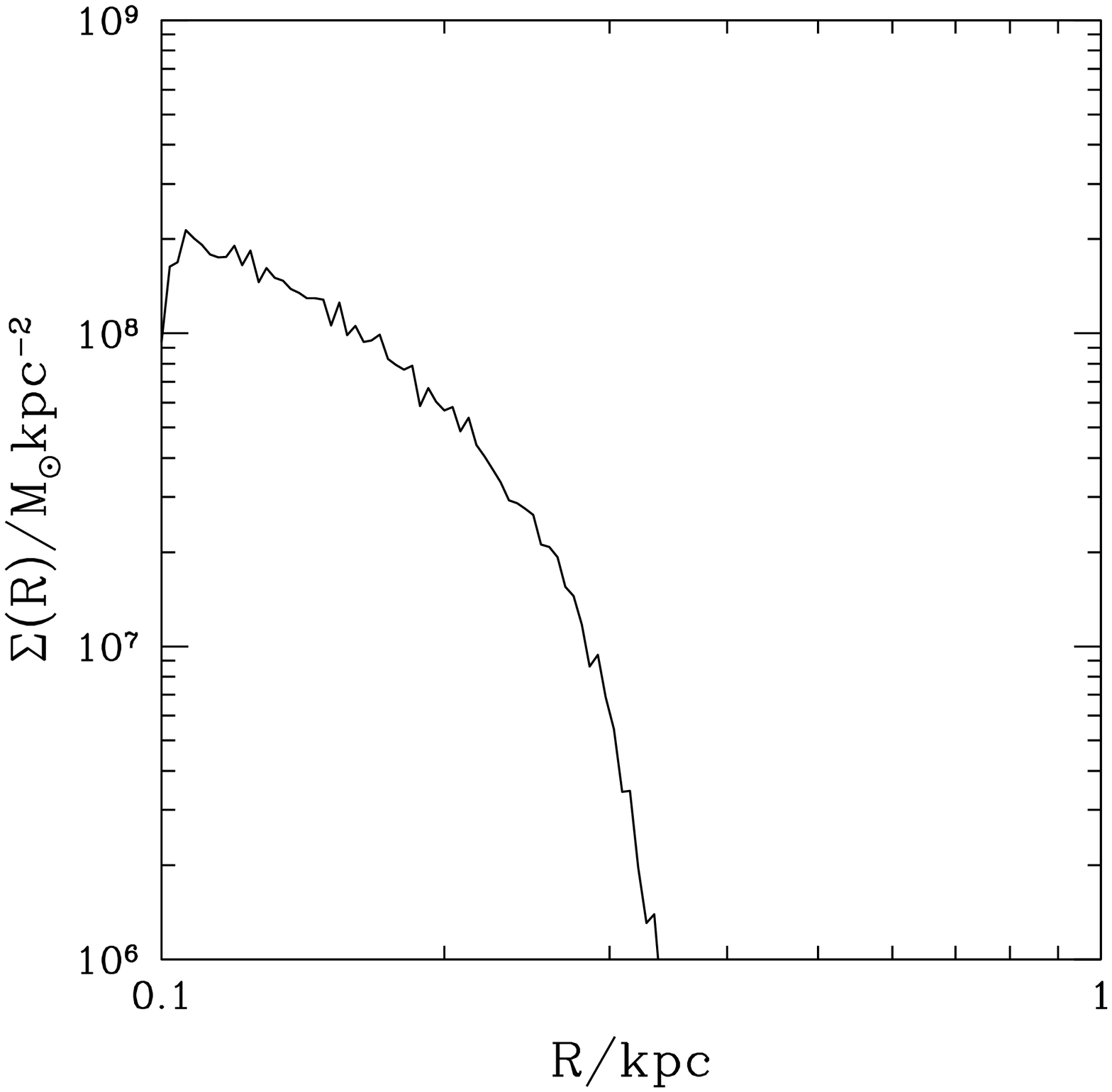,width=0.3\hsize}
\epsfig{file=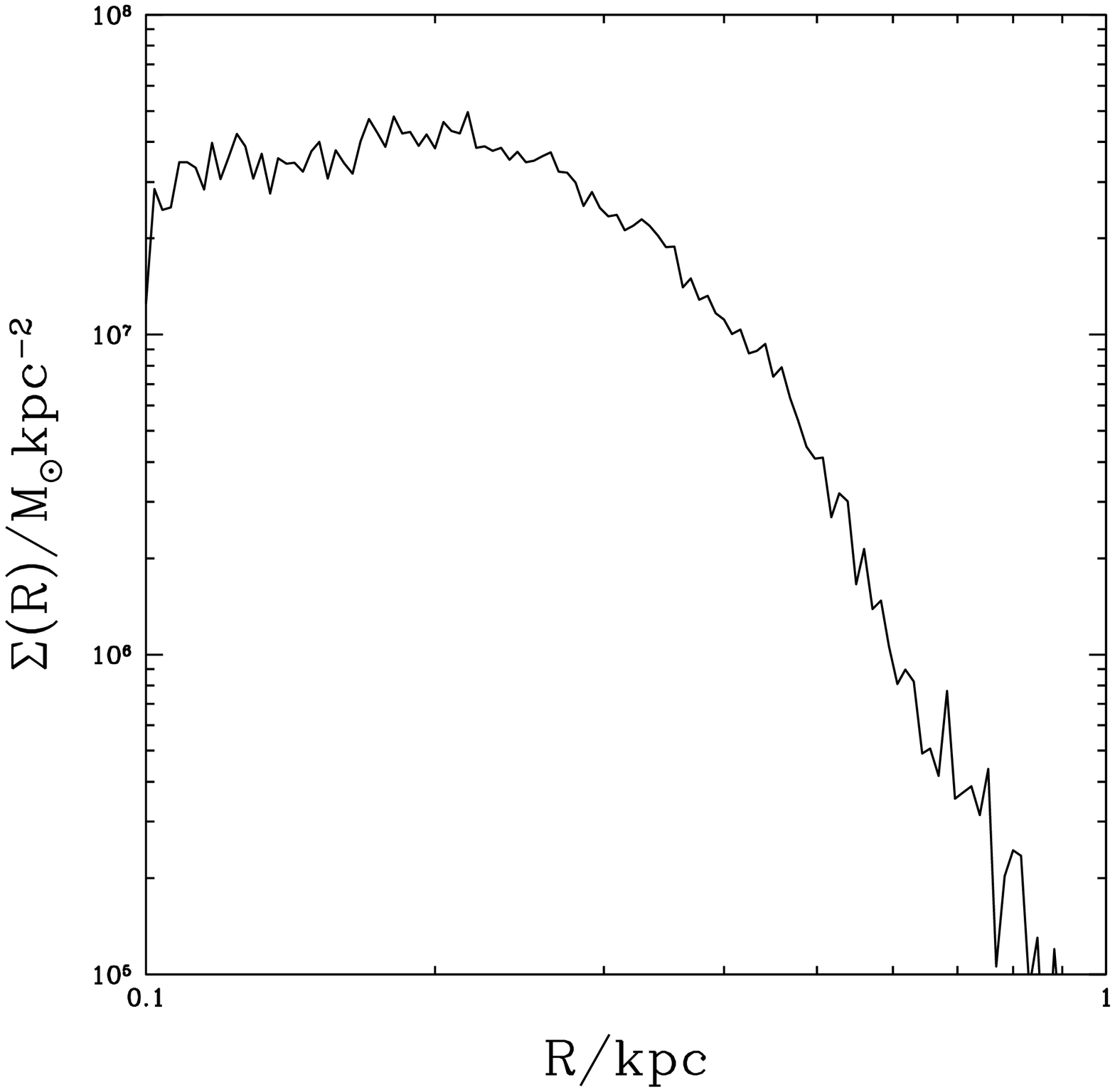,width=0.3\hsize}
}
\caption{A plot of surface density against radius for the disc formed
when $bV_0$ is $ 50 \, \textrm{kpc km s}^{-1} $ (left panel) and $ 100
\, \textrm{kpc km s}^{-1} $ (right panel).}
\label{correctring}
\end{figure*}

With these initial conditions, we find that a disc of similar size to
that observed in NGC~3379 can be reproduced if the specific angular
momentum of the incoming satellite, $bV_0$ is around $ 50 \,
\textrm{kpc km s}^{-1} $. The surface density profile of the disc
obtained using this value for the orbital angular momentum is shown in
Figure \ref{correctring}. In this situation, around 1\% of the gas
particles fall within the radius $10\pc$ and so are removed from the
calculation.

We conclude that in order to obtain a dust disc the approximate size
of the one observed it is necessary for the satellite galaxy to be on
an orbit which is almost precisely oriented towards the galactic
nucleus. In order to estimate of the likelihood of such an occurrence,
it is of course necessary to have some knowledge of the space and
velocity distribution of the prospective mergee population. This we do
not have. Numerical computations of galaxy formation through mergers
should in principle be able to give some insight into this
(e.g. Steinmetz 1999), but current simulations do not currently have
high enough mass resolution for our purposes.

We may, however, use the above results to estimate the fraction of
those merging satellites disrupted on their first pericentre passage
which give rise to a dust disc of the kind observed in NGC~3379. If we
assume that the satellite galaxy, as it falls into the main galaxy,
has a velocity approximately equal to that of the dispersion velocity
in the main galaxy, which is about 200 km s$^{-1}$, then for this
satellite give rise to the observed gas disc, we have seen that we
require $bV_0 \simeq 50 \, \textrm{kpc km s}^{-1} $, and thus that the
pericentre of the orbit to be $b \simeq 250 \pc$. For the satellite to
be disrupted on the first pass through the main galaxy, we have seen
(Section~\ref{tidalstripping}) that we require the pericentre distance
to be less than around $b \simeq 1 \kpc$. If we assume that incoming
satellites have their impact parameters evenly distributed over this
range then this would imply that only of the order of $4 \%$ of those
satellites disrupted on the first pass are on appropriate orbits to
form a disc such as observed in NGC~3379. Of course, the majority of
satellites have larger impact parameters. They would be captured by
the main galaxy, and would then slowly work their way inwards in the
manner described in Section~\ref{tidalstripping}, distributing their
gas and dust widely through the host galaxy. This should lead to
observable effects (Section~\ref{conclusions}). This in any case we
expect that for every merging satellite which could produce the
observed disc, there are many more which could not.

\section{Merging with a spiral galaxy}
\label{blackholes}

For the reasons discussed in the Introduction, we now extend the
foregoing analysis to consider the merging of a small satellite galaxy
with a spiral host galaxy. Our interest in this case centres not so
much on where the stripped gas may end up (since it likely in any case
to collide with and merge with the gas already in the gas disc of the
spiral), but rather on the fate of the nucleus of the satellite, or
more especially, on the fate of the black hole presumably contained
within it.

\subsection{Galactic models}

\subsubsection{The spiral}

We choose to use a simple model of a typical disc galaxy, in which the
galaxy is modelled using four axisymmetric components (representing
the central black hole/star cluster, the bulge, the disc and the dark
halo), each of which is represented by a Miyamoto-Nagai potential
(Binney and Tremaine 1987, Equation 2-50a), which takes the form

\begin{equation}
\Phi(R,z) = - \frac{ G M }{ \sqrt{ R^2 + \left( a + \sqrt{ z^2 + b^2 }
\right)^2 } },
\end{equation}
where we use a cylindrical coordinate system $(R,\phi,z)$.

The density due to this potential is

\begin{equation}
\rho(R,z) = - \frac{ b^2 M }{ 4 \pi } \frac{ a R^2 + \left( a + 3 \sqrt{
b^2 + z^2 } \right) \left( a + \sqrt{ b^2 + z^2 } \right)^2 }{ \left( R^2
+ \left( a + \sqrt{ b^2 + z^2 } \right)^2 \right)^{5/2} \left( b^2 +
z^2 \right)^{3/2} }.
\end{equation}

We note that this reduces to a Plummer sphere in the case of $a =
0$. From these expressions, and using the assumption that the internal
galactic dynamics consists of an isotropic velocity dispersion coupled
with a mean azimuthal streaming motion, we can calculate the velocity
dispersion and mean streaming velocity of each component, via (Binney
and Tremaine, 1987, Equations 4-65 and 4-66):

\begin{eqnarray}
\sigma^2(R,z) & = & \frac{1}{\rho(R,z)} \int_z^\infty \rho(R,z)
\frac{\partial \Phi(R,z')}{\partial z'} \textrm{d} z' \\
\overline{v}_\phi^2(R,z) & = & R \frac{\partial \Phi(R,z)}{\partial R}
\nonumber\\ & + & \frac{R}{\rho(R,z)} \frac{\partial}{\partial R}
\int_z^\infty \rho(R,z') \frac{\partial \Phi(R,z')}{\partial z'}
\textrm{d} z'
\end{eqnarray}

As before, enlarging the parameter space to make more complicated
assumptions about the stellar distribution function is not warranted
given the other simplifications we adopt here.  In any case, the
results are not likely to be strongly affected unless the stellar
distribution function is extremely anisotropic.

For our disc galaxy, we take the parameters used by Sofue (1996) for
the Milky Way which assumes: (i) a spherically symmetric ($a=0$)
nuclear star cluster of mass $5 \times 10^9 \ \msun$, and scale $b =
120\pc$ (ii) a spherically symmetric bulge of mass $10^{10} \msun$
and scale $b = 750\pc$; (iii) an axisymmetric disc of mass $1.6 \times
10^{11} \msun$, radial scale $a = 6\kpc$ and vertical scale $b = 500
\pc$; and (iv) a spherically symmetric dark halo of mass $3 \times
10^{11} \msun$ and scale $b = 15\kpc$.

\subsubsection{The satellite}

As above, we model the satellite galaxy in this case as a Plummer
sphere, subject to tidal stripping but with the addition of a central
point mass to represent a nuclear black hole and surrounding star
cluster. We increase the mass of the satellite to be $4 \times 10^9
\msun$ but retain the same scale length of $b = 0.4 \kpc$. This is
consistent with the models of Sagittarius of Helmi and White (2001).
This represents the inclusion of dark matter in the model for the
dwarf.  This could have be added into the previous simulations, but
would have had little effect as the satellite was very rapidly
destroyed by tidal stripping.  We set the central point mass to be $1
\times 10^7 \msun$, although we shall see later that our simulations
are basically independent of this.

\subsection{Orbital dynamics}
\label{blackholedynamics}

We model the orbital dynamics of the satellite in a manner using
basically the same ideas as used above
(Section~\ref{tidalstripping}). The satellite galaxy is subject to the
gravitational influence of the main galaxy, to dynamical friction and to
tidal stripping.

The gravitational force due to the main galaxy is straightforward to
calculate. Due to the axisymmetry of our model, there are only two
independent components to consider. The radial force component (in a
cylindrical sense, that parallel to the plane of the disc) is

\begin{equation}
F_R = 
- \frac{\partial \Phi}{\partial R} =
- \frac{ G M R }{ \left( R^2 + \left(
a + \sqrt{ b^2 + z^2 } \right)^2 \right)^{3/2} }
\end{equation}
and the vertical component (perpendicular to the disc),

\begin{equation}
F_z = 
- \frac{\partial \Phi}{\partial z} = 
- \frac{ G M z \left( a + \sqrt{ b^2
+ z^2 } \right) }{ \left( R^2 + \left( a + \sqrt{ b^2 + z^2 }
\right)^2 \right)^{3/2} \left( a + \sqrt{ b^2 + z^2 } \right)}.
\end{equation}

The force in both directions is calculated for each of the four
components of our model galaxy and these forces summed to produce the
resultant force on the satellite due to the main galaxy.

To calculate the dynamical friction on the satellite, we first
calculate the mean streaming velocity, $\overline{v}_\phi$, of the
disc component of the galaxy at the satellite's current location using
Equation~16 (note that all other components have zero mean streaming
velocity since they are spherically symmetric) and from this we
calculate the relative velocity of the satellite which respect to the
disc. Second we compute the local velocity dispersion of each
component of the galaxy. Then, finally, we use the standard
gravitational drag formula (Equation~\ref{chandra}) to calculate the
dynamical friction force on the satellite. Once again, the
contribution from all four components of the galaxy is then summed,
and this added to the gravitational force to produce the total force
on the satellite.

After each timestep in our integration, we apply simple tidal stripping
to the satellite. We calculate the mass internal to the satellite's
current position, via

\begin{equation}
M(R,z) = 2 \pi \int_0^{\sqrt{R^2+z^2}} \int_0^\pi \rho(r,z) r^2 \sin \theta
\textrm{d} \theta \textrm{d} r
\end{equation}
and then use this mass to calculate the current tidal radius of the
satellite, as given in Equation~\ref{tidalradius}. Whilst this is not
strictly accurate, it is an adequate approximation for our current
purposes.  As we shall see below, the stripping of the satellite is
not the most important factor in these simulations.

Again, we use a variant of the `\texttt{odeint}' routine from
Numerical Recipes \shortcite{numrec} to control the timestep in the
simulation, so that we use small timesteps when necessary, but larger
ones when possible to speed up our simulation.

\subsection{Results}
\label{bh-results}

Since we know from our previous set of calculations that the only
satellite trajectories which are likely to have a chance of reaching
the centre of the host galaxy are those which are almost radial, we
assume that the satellite galaxy is initially approaching the host
galaxy from a large distance ($\geq 100\kpc$).

To be specific, we define the centre of the Seyfert as the origin of
our coordinate system, and then consider the trajectories of small,
incoming galaxies which start on the plane which has a normal vector
at an angle $90 - \alpha$ from the axis of rotation of the Seyfert and
has a point of closest approach of $100 \kpc$ to the origin. We
consider satellite orbits which start in this plane, and which have
their initial velocity being perpendicular to the plane. We shall call
the angle $\alpha$ the `approach angle', as it corresponds to the
angle between the initial velocity vector and the plane of the host
galaxy. Thus $\alpha = 90\degr$ corresponds to an orbit which is
initially directed at right angles to the disc of the host galaxy.
For this paper, we investigated orbits with approach angles of
$0\degr$, $15\degr$, $30\degr$, $45\degr$, $60\degr$, $75\degr$,
$85\degr$, and $90\degr$ and with two sets of initial velocities: $100
\kms$ and $200 \kms$.

We continue to track the satellite's orbit either until the satellite
has settled into the plane of the disc, after which orbital evolution
is very slow, or until the satellite reaches a radius,
$r_{\rm sph}$ at which the potential of the host galaxy is
spherical and so there will be no further changes in the direction of
satellite's orbital angular momentum. For our model, we take
$r_{\rm sph}$ to be $200 \ \pc$.

\begin{figure}
\centerline{
\epsfig{file=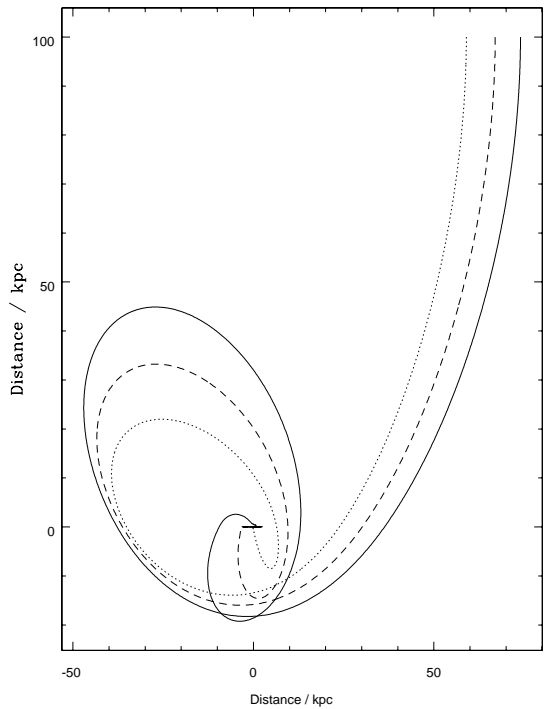,width=0.9\hsize}
}
\caption{Three orbits in the disc galaxy; the disc is perpendicular to
the page. The approach angle, $\alpha$ is $90\degr$ and the initial
velocity is $100 \kms$ in all three cases.}
\label{bhorbit}
\end{figure}

The main outcome of our orbital computations is that most satellites
settle fairly quickly into the orbital plane of the disc galaxy. This
comes about simply because the gravitational drag due to the plane is
relatively large because the dispersion velocity of the disc component
is relatively small. The fraction of the initial conditions which
result in the satellite reaching the central regions of the main
galaxy \emph{before} it settles into the plane of the disc is
relatively small, but not vanishingly so. Those that do reach the
centre are generally those for which their first dynamically
significant impact with the disc is in the centre of the galaxy. Three
orbits, chosen such that two of them (the solid and dotted lines) are
those which reach the centre of the galaxy, are shown in
Figure~\ref{bhorbit}. Once the satellite has settled into the plane
its orbital evolution becomes very slow, and does not in general reach
the centre within a Hubble time.  We should note that for our
approximations of a static potential for the disc to be valid, we
require that the satellite not be more massive than the mass it
interacts with when passing through the disc.  The satellite will
interact with material in a cylinder of radius of order of $G M /
v^2$, where $v$ is the velocity of the satellite when it passes
through the disc.  This mass is always greater than the current mass
of the satellite when passing through the disc, except for the
earliest passes through the disc when there is no significant drag.

For each orbit which reached $r_{\rm sph}$ before settling into the
plane of the disc, (i.e. which reached the centre in less than a
Hubble time) we recorded the angle, $\beta$, which the satellite's
final orbital angular momentum vector made with the normal to the
plane of the disc.  Thus $\beta = 0\degr$ means the satellite's
angular momentum is parallel to that of the rotational angular
momentum of the disc, i.e the final orbit of the satellite lies in the
plane of the galaxy disc.

\begin{figure}
\centerline{\epsfig{file=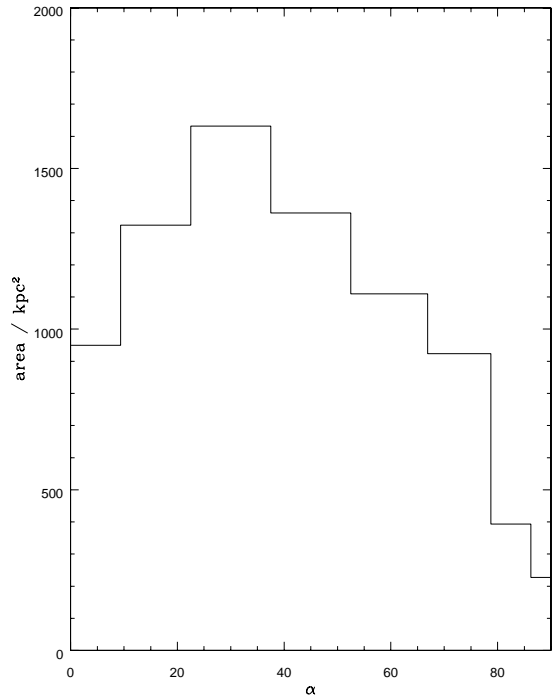,width=0.9\hsize}}
\caption{The contribution to the cross-sectional area for the
satellite reaching $r_{\rm sph}$ before settling into the disc of the
host galaxy for various approach angles $\alpha$ (see
Section~\ref{bh-results}).}
\label{centrearea}
\end{figure}

In general, the plane ($\alpha = const.$), from which we launch our
initial trajectories contains a small number of regions (typically
around 5) from which the satellite reaches the centre of the host
galaxy, but these are spread out over a relatively large area. To find
the cross-sectional area $A \left( \alpha \right)$ at each approach
angle, we first undertook a low-resolution scan (out to a radius where
satellite simply flew past the galaxy), to find the regions in which
the satellite reached the centre before reaching the plane of the
galaxy. We picked out the areas of interest by eye, and then
investigated those regions at higher resolution. The cross-sectional
area was then calculated simply by summing the area of these regions.
The exception to this is the face-on case, which is rotationally
symmetric in the plane of the disc. Here we simply undertook a high
resolution scan along one line parallel to the plane of the disc and
then calculated the cross-sectional area from the circular annuli
produced. In Figure~\ref{centrearea}, we plot the contribution
$A\left(\alpha\right) \cos \alpha$ to the total cross-sectional area
made by each approach angle $\alpha$. The factor of $\cos \alpha$
accounts for the fact that we expect fewer satellites to have approach
angles close to $90\degr$, assuming the incoming flux of satellites is
isotropic.

\begin{figure}
\centerline{\epsfig{file=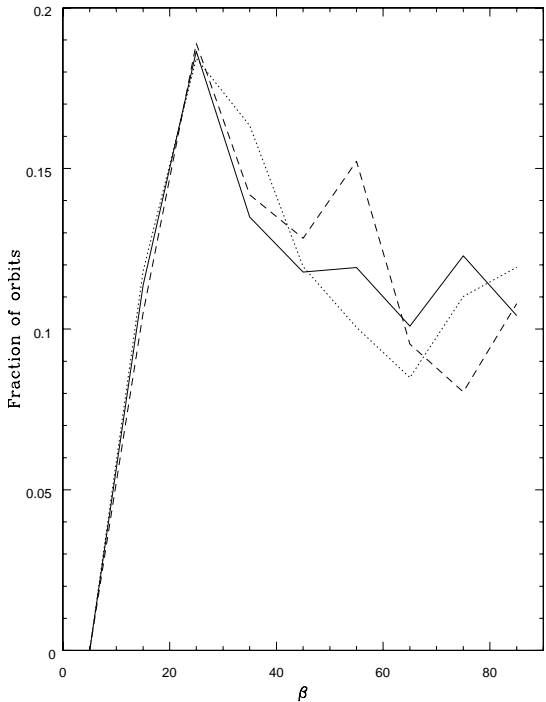,width=0.9\hsize}}
\caption{The distribution of the angle $\beta$ between the angular
momentum vectors of the infalling satellite and the central black hole
of the Seyfert galaxy for different approach angles $\alpha$ of the
satellite's initial orbit: 0\degr (dashed), 30\degr (dotted) and
60\degr (solid).}
\label{fig-angles}
\end{figure}

We have also determined the angle $\beta$ between the orbital angular
momentum of the satellite's orbit when it reaches the centre ($r <
r_{\rm sph}$) and the normal to the plane of the galaxy disc.
Figure~\ref{fig-angles} shows the distribution of $\beta$ for
different approach angles $\alpha$. We note that there is little
change with $\alpha$.  The absence of orbits which end with a low
value of $\beta$ is due to the fact that these orbits are close to the
plane of the disc, and so quickly settle into the plane of the disc
and are never reach the radius $r_{\rm sph}$.

\begin{figure}
\centerline{\epsfig{file=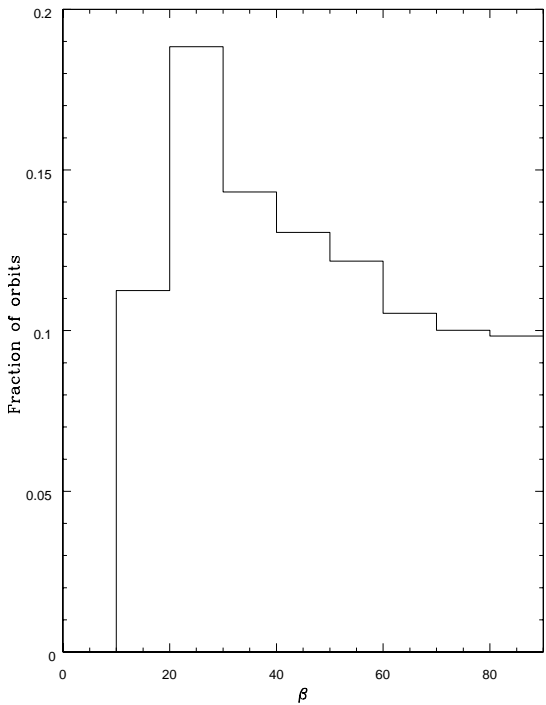,width=0.9\hsize}}
\caption{The overall distribution of the angle $\beta$ assuming that
the initial distribution of orbital inclinations is uniform.}
\label{fig-angles2}
\end{figure}

We also compute an overall distribution for the angle $\beta$ for
those mergees which reach the centre of the host galaxy. In doing so
we encounter the problem that the distribution of the initial orbits
of the satellites' orbits is unknown. In the absence of any strong
evidence to the contrary, we assume that the incoming merging
satellites approach from random directions. With this assumption, we
have computed the distribution for $\beta$, and this is as shown in
Figure~\ref{fig-angles2}. We note that this distribution is reasonably
uniform in $\beta$, except for a zone of avoidance for angles $\beta
\lta 20 \degr$.

We have repeated the above experiments with the black hole in the
centre of the dwarf galaxy having a mass from $10^5 \msun$ to $10^8
\msun$ (as opposed to the $10^7 \msun$ used previously), and obtain no
significant differences. This may have been expected from the fact
that the satellite is not stripped down to a small mass by the time it
reaches a radius of $200 \pc$ from the centre of the Seyfert.

\subsection{Comparision with observations}
\label{observations}

A comparision of our model with the available observational data is
not entirely trivial. In order to obtain a quantitative result, we
follow the method of Kinney et al (2000, and references therein). A
brief description of the method is given here.

Two parameters are observed for every galaxy used in Kinney et al.:
$i$, the inclination of the galaxy's disc to the plane of the sky, and
$\delta$, the difference between the position angles of the galaxy's
major axis and that of the jet when projected onto the plane of the
sky.  Each $\left( i, \delta \right)$ pair constrains the jet to lie
on a pair of great circles on a sphere about the nucleus of the
galaxy.  The angle $\beta$ between the jet and the rotation axis of
the disc varies around these great circles.  We assume that the chance
of a jet being at any point on either of these great circles is given
by the probability of our simulations producing the value of $\beta$
(as shown in Figure~\ref{fig-angles2}) for that point on the great
circle. At a fixed inclination, different values of $\delta$ give
great circles which have different orientations in space, and hence we
calculate the probability of obtaining a given value of delta given
the inclination, $p\left( \delta | i \right)$. We then integrate this
distribution from $\delta = 0$ to $\delta = \delta_k$ where $\delta_k$
is the observed value of $\delta$ for the galaxy $k$ to obtain the
centile point $c_k$. The distribution of centile points across the
data set should then be uniform over the range
$\left[0,1\right]$. Full details of this technique can be found in
Kinney et al. (2000).

We perform a K-S test to compare our model with the the $60 \,
\mu\textrm{m}$ data of Kinney et al. (2000). We find that the data is
consistent with being drawn from our model at the 97\% level.

\section{Discussion}
\label{conclusions}

We have undertaken an investigation into the possibility that the
activity of, and jet direction in, at least some active galactic
nuclei might be fuelled by the infall and merger of small satellite
galaxies. 

\subsection{Cross-sections for mergers}

We considered first the evidence that many of the host
galaxies active nuclei in the 3CR catalogue are found to have discs of
dust around their nuclei. Taking this as {\it prima facie} evidence
that the nuclear activity might be related to infall, we have modelled
the disruption of a small satellite galaxy by a larger elliptical,
taking the particular case of NGC~3379, which while not in the 3CR
catalogue, does show a radio bright nucleus and does display a dust
disc.

Our modelling procedure is of necessity somewhat crude. We have
adopted a realistic model for the inner regions of the host elliptical
where the disruption of the infalling satellite takes place, but have
not taken proper account of a possible massive, but low density, dark
halo. We considered idealised models for the internal structure of the
satellite, one with and one without a central density cusp, but found
that the results did not depend significantly upon which models we
used. Gravitational drag on the orbit of the satellite was modelled
approximately using the standard Chandrasekhar formula, suitably
parameterized by comparison with full numerical computations taken
from the literature.

Probably the least realistic part of our modelling procedure is the
description of the disruption of the satellite and the subsequent
dynamics of the gas it contained. For example, we have assumed that
the main galaxy is sufficiently devoid of gas that ram pressure
stripping of gas from the satellite can be ignored, and that the
effect of the interactions between the stripped gas and the gas
already present may be ignored. This may not always be the case. Even
so, because our description does contain much of the relevant physics
it is unlikely that our results will be grossly in error. The
infalling satellite is disrupted at the point where tidal effects of
the host come into play using simple truncation of the underlying
model. This ignores the dynamical response of the satellite to the
truncation process, and so, if anything, tends to overestimate the
radius at which disruption occurs. On the other hand, we have
implicitly assumed that the gas and stars in the satellite occupy the
same volume, and thus strip gas particles in proportion to the mass
removed. If the gas were more centrally located than the bulk of the
mass, then our procedure would underestimate the radius at which the
gas from the satellite is deposited. However, as can be seen from
Figure \ref{strip}, the radius at which tidal disruption takes place
is fairly well-defined, with the radius by which 90 per cent of the
satellite has been tidally removed being almost independent of the
model for the internal structure. The only way to bring about a
significant change in the radius at which the gas is deposited is to
use the cusped model and to assume that all the gas lies within the
radius occupied by the central one percent of the total galaxy
mass. This would imply for a gas to mass ratio of a percent that the
gas was dynamically important in the nuclear region of the
satellite. The modelling of the dynamics of the gas once it has been
stripped conserves mass and angular momentum, but uses a simple
algorithm to allow intersecting streams of gas to shock and lose
energy by radiation. This is probably an adequate approximation for
the current paper.

Despite the relative crudeness of our modelling procedure, we are able
to draw some conclusions which we believe are relatively robust. We
find that while it is possible to reproduce the dust disc structure
seen in NGC~3379 via the infall of a satellite galaxy, it seems likely
that this type of event is very rare. Our initial simple model for the
satellite, which started within the galactic halo on an orbit with a
given eccentricity, was not able to reproduce the structure seen in
NGC~3379. A more realistic model, in which the satellite starts well
outside the galaxy with a small impact parameter was found to be
capable of reproducing the observed structure, provided that the
satellite galaxy has an initial angular momentum about the galactic
nucleus of around $ 50 \, \textrm{kpc km s}^{-1} $.  Given the lack of
observational or theoretical evidence as to the distribution of the
orbits of dwarf satellites, we cannot make a quantitative statement
about how likely this is, but we do not believe that this event is
common.  If we assume a typical velocity of around $200\kms$ for the
satellite when at infinity, this means the initial orbit must have an
impact parameter of the order of $250\pc$, which would represent a
small fraction of phase space if the initial orbits were uniformly
distributed.  Even if the orbital distribution were moderately biased
towards low angular momentum orbits, this would still imply that the
formation of a structure similar to that in NGC~3379 being formed by
this method is a rare event.

With regard to powering a single AGN, this model produces results
which are not inconsistent with current theories. To order of
magnitude level, if an event forming a disc similar to that observed
in NGC~3379 occurs, we find that a few percent of the gas particles
reach a distance of $10 \pc$ from the nucleus. At this point it is
reasonable to assume that they will be swallowed by the central black
hole within a reasonably short time span \cite{shlosman90,bekki00}. If
we assume that a few percent of the mass of the original dwarf galaxy
was gas, this gives around $10^6 \ \msun$ being swallowed by the black
hole, which, at an accretion rate of $0.1 \,
\textrm{M}_\odot/\textrm{yr}$, will power a low level AGN for around
$10^7 \yr$. This is not unreasonable.

However, if we further assume that all galaxy activity is triggered
only by events of the above type, then the model becomes harder to
sustain. If we assume that every merger triggers nuclear activity
which lasts around $10^7$ years, and if around $1\%$ of elliptical
galaxies are active, then this implies that each elliptical galaxy has
undergone about 10 such events over the lifetime of the
universe. However, if only a few percent of mergers occur at
sufficiently low angular momentum to give rise to nuclear activity,
then the total number of satellites swallowed by the host over its
lifetime rises to around 300 to 1000. If each satellite has a mass of
around $10^9 \msun$ as we assume here, this implies the addition of a
substantial amount of mass. Even if our estimates are too pessimistic
with regard to the efficiency with which merger process is able to
fuel the nucleus, it is clear that we would nevertheless expect that
essentially all elliptical galaxies should show some evidence of such
a merger having occurred, and perhaps more than one. In this regard,
we note that whilst van Dokkum and Franx \shortcite{vandokkum95} found
dust structures in approximately half of their sample, more recent
more sensitive surveys \cite{pastoriza00} are finding dust in a higher
proportion of elliptical galaxies, including some in which dust
remained undetected in the van Dokkum and Franx survey.

\subsection{Directionality}

Having established that the likelihood of any particular merging
satellite galaxy reaching the host nucleus is relatively small, we
then turn our attention to incoming orbits for which the successful
mergees actually reach the nucleus. The motivation for this part of
the investigation is the discovery (Kinney et al. 2000) that the jet
directions in the centres of Seyfert galaxies appear to bear no
relation to the orientation of the discs of the host galaxies. The
possibility we are investigating here is whether the mergees reaching
the centre do so with an angular momentum distribution which could
produce the observational data observed by Kinney et al.

For this case we have taken the host galaxy to be a spiral galaxy, and
have employed the same procedures for orbital decay and tidal
stripping as before. Again we find that the cross-sectional areas for
merging of the satellite with the host nucleus are relatively small,
but we now concentrate our analysis on the angular momentum of the
orbits of those satellites (or what is left of them) when they reach
the nuclear regions of the host. Those satellite nuclei which do reach
the centre in a finite time (that is, before settling into the disc)
have the angle between their angular momentum vector and the disc
normal ($\beta$) distributed approximately randomly, as is shown in
Figure~\ref{fig-angles2}.  The deficit at low values of $\beta$ comes
about because these satellites are close to, or in, the disc of the
Seyfert galaxy, and as such rapidly settle down into the plane.  We
have noted that is distribution is in good agreement with the data
presented in Kinney et al. (2000).
 
The simulations we have performed here give us a distribution for the
angle $\beta$. However, in order to relate these results to the
observed jet distribution, we need the resultant spin of the black
hole formed between the merger of the two holes. For the general case,
this is an unsolved problem in general relativity and hence we again
adopt a simple prescription. We assume the mass of the dwarf's black
hole, $M_D$, is much less than the mass of the Seyfert's black hole,
$M_{\rm AGN}$ and can then use the prescription from Colbert and
Wilson (1995) to first order to obtain the spin of the daughter of the
merger as

\begin{equation}
L_S = \frac{ 2 \sqrt{3} G M_D M_{\rm AGN} }{ c }
\end{equation}
where $L_S$ is the spin of the resultant hole.  This applies only for
a merger between two black holes which were not originally spinning,
which we note is not the case here. If $M_D/M_{\rm AGN} = 0.1$, then
the resultant hole is a Kerr black hole with spin parameter $a=0.35$,
where $a$ is the angular momentum of the hole in units of $GM^2/c$ and
lies between 0 and 1. Thus the accumulation of a few satellite nuclear
black holes by the black hole in the AGN nucleus, from orbital
directions which are more of less randomly distributed can lead to
random orientations for the spins of the AGN nuclear black holes. A
more detailed discussion of this is given by Merritt (2002).

\section{Conclusions}

We have investigated the dynamics of the merging process in the
minor merger hypothesis for active galactic nuclei. Our analysis has
resulted in two main conclusions.

First, for the satellite galaxy to be able to merge directly with the
nucleus of the host galaxy (for example, to give rise to the compact
dust discs which are seen in early type active galaxies) requires the
initial orbit of the satellite to be well aimed. The corollary of this
is that for each merging satellite which gives rise to a compact
nuclear dust disc, there must be many which do not. Similarly, in disc
galaxies, for every minor merger which causes an active nucleus, there
must be many mergers which simply give rise to a merger of the
incoming satellite with the galaxy disc.

Second, in the case of the host galaxy being a disc galaxy, if the
initial orbits of the satellites are randomly oriented with respect to
the host galaxy, then the orbits of those which reach the host nuclear
regions in a reasonable time, are also fairly randomly oriented once
they reach the nucleus. We note that this result might be able to
provide an explanation of why the jet directions in the nuclei of
Seyfert galaxies are apparently unrelated to the plane of the galaxy
disc. The distribution of angles $\beta$ between the final orbit of
the successful merger candidates and the Seyfert disc which we find
once we have averaged over an assumed isotropic incoming satellite
distribution is consistent with the observational data presented by
Kinney et al (2000).

\section{Acknowledgments}

We thank Frank van den Bosch for providing us with the diagrams in
Figure~\ref{fig-vandenbosch}. We would also like to thank an anonymous
referee for constructive comments made on earlier versions of this
paper.

\label{lastpage}


\begin{thebibliography}{}
\bibitem[\protect\citename{Bekki }2000]{bekki00} Bekki~K., 2000, ApJ,
545, 753--757.
\bibitem[\protect\citename{Binney \& Tremaine }1987]{binney87}
Binney~J.J., Tremaine~S., 1987, Galactic Dynamics, Princeton
University Press.
\bibitem[\protect\citename{Chatzichristou }2001a]{chris00a}
Chatzichristou~E.T., 2000, ApJS, 131, 71--94.
\bibitem[\protect\citename{Chatzichristou }2001b]{chris00b}
Chatzichristou~E.T., 2000, ApJ, 544, 712--733.
\bibitem[\protect\citename{Chatzichristou }2001a]{chris01a}
Chatzichristou~E.T., 2001, ApJ, 556, 653--675.
\bibitem[\protect\citename{Chatzichristou }2001b]{chris01b}
Chatzichristou~E.T., 2001, ApJ, 556, 676--693.
\bibitem[\protect\citename{de Koff et al.} 2000]{dekoff00} de Koff~S.,
2000, ApJS, 129, 33--59.
\bibitem[\protect\citename{Ferrari }1999]{ferrari99} Ferrari~F.,
Pastoriza~M.G., Macchetto~F., Caon~N., 1999, A\&AS, 136, 269--284.
\bibitem[\protect\citename{Gebhart et al. }2000]{gebhardt00}
Gebhardt~K. et al., 2000, AJ, 119, 1157--1171.
\bibitem[\protect\citename{Helmi \& White }2000]{helmi00} Helmi~A.,
White~S.D.M., 2001, MNRAS, 323, 529--536.
\bibitem[\protect\citename{Hernquist \& Weinberg }1989]{hernquist89}
Hernquist~L., Weinberg~M.D., 1989, MNRAS, 238, 407--416.
\bibitem[\protect\citename{Kinney et al. }2000]{megapaper}
Kinney~A.L., Schmitt~H.R., Clarke~C.J., Pringle~J.E., Ulvestad~J.S.,
Antonucci~R.R.J., 2000, ApJ, 537, 152--177.
\bibitem[\protect\citename{Lima~Neto et al. }1999]{limaneto99}Lima~Neto~G.B.,
Gerbal~D., M\'arquez~I., 1999, MNRAS, 309, 481--495.
\bibitem[\protect\citename{Lin \& Pringle }1976]{lin76} Lin~D.N.C.,
Pringle~J.E., 1976, IAU Symposium 73, 237--253.
\bibitem[\protect\citename{Magorrian et al. }1998]{magog98} Magorrian~J.,
et al. 1998, AJ, 115, 2285
\bibitem[\protect\citename{Martel }1999]{martel99} Martel~A.R.,
Baum~S.A., Sparks~W.B., Biretta~J.A., Verdoes Kleijn~G.2000,
ApJS, 130, 267--338.
\bibitem[\protect\citename{Merritt }2002]{merrit02} Merritt~D., 2002,
ApJ, 568, 998--1003.
\bibitem[\protect\citename{Pastoriza et al. }2000]{pastoriza00}
Pastoriza~M.G., Winge~C., Ferrari~F., Duccio~Macchetto~F., Caon~N.,
2000, ApJ, 529, 866--874.
\bibitem[\protect\citename{Press et al. }1992]{numrec} Press~W.H.,
Teukolsky~S.A., Vetterling~W.T., Flannery~B.P., 1992, Numerical
Recipies in C: The Art of Scientific Computing: Second Edition, CUP
\bibitem[\protect\citename{Schmitt et al. }2002]{schmitt02}
Schmitt~H.R., Pringle~J.E., Clarke~C.J., Kinney~A.L., ApJ, submitted.
\bibitem[\protect\citename{Scheuer \& Feiler }1996]{scheuer96}
Scheuer~P.A.G., MNRAS, 282, 291.
\bibitem[\protect\citename{Steinmetz }1999]{steinmetz99}
Steinmetz~M., Ap\&SS, 269/270, 513--532.
\bibitem[\protect\citename{Shlosman et al. }1990]{shlosman90}
Shlosman~I., Begelman~M.C., Frank~J., 1990, Nat, 345, 679--686.
\bibitem[\protect\citename{Sofue }1996]{sofue96} Sofue~Y., 1996, ApJ,
458, 120.
\bibitem[\protect\citename{Taniguchi }1999]{taniguchi99} Taniguchi~Y.,
1999, 524, 1, 65--70.
\bibitem[\protect\citename{Taylor \& Babul }2001]{taylor01}
Taylor~J.E., Babul~A., 2001, ApJ, 559, 716--735.
\bibitem[\protect\citename{van den Bosch et al. }1999]{vandenbosch99}
van~den~Bosch~F.C., Lewis~G.F, Lake~G., Stadel~J., 1999, ApJ, 515,
50--68.
\bibitem[\protect\citename{van Dokkum \& Franx }1995]{vandokkum95}
van~Dokkum~P.G., Franx~M., 1995, AJ, 110, 2027--2036.
\bibitem[\protect\citename{Vel\'azquez \& White }1999]{velazquez99}
Vel\'azquez~H., White~S.D.M., 1999, MNRAS, 304, 254--270.
\bibitem[\protect\citename{Verdoes~Kleijn et al. }2000]{verdoes00}
Verdoes Kleijn G.A. et al., 2000, in Schilizzi R., Vogel S., Paresce
F., Elvis M., eds., Galaxies and their Constituents at the Highest
Angular Resolutions, Manchester UK, August 2000
\bibitem[\protect\citename{Wilson \& Colbert }1995]{wilson95}
Wilson~A.S., Colbert~E.J.M., 1995, ApJ, 438, 62--71.
\bibitem[\protect\citename{Zaritsky \& White }1988]{zaritsky88}
Zaritsky~D., White~S.D.M., 1988, MNRAS, 235, 289--296.
\end{thebibliography}
\end{document}